# All-optical polarization control and routing by nonlinear interferometry at the nanoscale

*Yigong Luan,[1] Attilio Zilli,[1] Agostino Di Francescantonio,[1] Vincent Vinel,[2] Paolo Biagioni,[1] Lamberto Duò,[1] Aristide Lemaître,[3] Giuseppe Leo,[2] Michele Celebrano[1\*], Marco Finazzi[1\*]*

1. Department of Physics, Politecnico di Milano, Milano 20133, Italy

2. Université de Paris, CNRS, Laboratoire Matériaux et Phénomènes Quantiques, Paris 75013, France

3. Centre de Nanosciences et de Nanotechnologies, CNRS, Université Paris-Saclay, Palaiseau 91120, France

[\*]Correspondence to: michele.celebrano@polimi.it; marco.finazzi@polimi.it

Optical metasurfaces are rapidly establishing as key-enabling platforms for nanophotonics applications. Along with the ability of taming light in subwavelength thicknesses, they can feature multiple functionalities in one device. The generation and control of light polarization by metasurfaces already provided a route towards ultracompact polarimetry devices in the linear regime. If translated to the nonlinear optical regime it may become a key-enabling tool in nonlinear imaging, optical holography and sensing. Here, we report the experimental ultrafast all-optical polarization modulation of upconverted light by all-dielectric metasurfaces via nonlinear interferometry. By controlling the relative phase between a pump beam at $\omega$ and its frequency-double replica at $2\omega$, we can set the phase relation between two frequency-degenerate upconverted processes at $3\omega$ – Sum-Frequency Generation (SFG) and Third-Harmonic Generation (THG) – stemming from an AlGaAs metasurface. By exploiting the opposite parity of the two nonlinear process and tuning their relative powers, we can achieve modulation of the polarization states of the light emitted at $3\omega$ between linear and circular states with a degree of circular polarization (DOCP) up to 83%. In particular, circularly polarized light features opposite handedness symmetrically located in the Fourier space, at coincidence with the first diffraction orders of the metasurface. Moreover, by toggling the phase delay by π, the handedness can be fully switched between the diffraction. Our work adds an additional layer of modulation in polarization beyond intensity to all-optical routing with precise phase control. The capability to route circular polarization states in the *k*-space holds significant potential for chiral sensing and advanced imaging techniques.

**INTRODUCTION**

Optical metasurfaces – engineered ultrathin, planar devices featuring subwavelength thickness – are revolutionizing light manipulation, offering several new degrees of freedom for light-matter interaction control[1]. Beside their compactness, compared to conventional bulk optical components, metasurfaces can be engineered to yield simultaneously multiple optical functionalities on a single platform[2], such as wavefront shaping[3,4], spectral[5,6] and temporal[7,8] control. The control and manipulation of light polarization[9,10] and parametric nonlinear light conversion[11–13] are among the key functionalities holding great potential for the development of advanced optical devices. In particular, the former has already found a notable application in polarimetric cameras for smart image recognition[14,15]. The cooperative behavior of the two above phenomena in optical metasurfaces was also recently exploited to realize all-optical switching[16], polarization imaging[17,18] and chiral sensing[19]. Reconfigurability represents a long-sought feature that can boost the application of metasurfaces to many fields in photonics, with particular impact to the rapidly evolving field of metasurface-based optical analog computing[20,21]. This has been thus far achieved in metasurfaces and nanostructures via electro-optical[22–24], thermo-optical[25,26], and all-optical[27–29] control with commutation frequencies exceeding the GHz. Among these approaches, the latter is particularly promising since it combines moderately high commutation efficiencies with the potential to achieve extremely high computational frequencies. The experimental realization of all-optical control of light by metasurfaces was thoroughly investigated both in the linear[27,28] and nonlinear[30,25,31] regime, where large modulation efficiencies can be achieved[25]. Although most of these studies focused on intensity modulation, it is known that resorting to phase modulation may enable THz-speed reconfigurability[32]. In this framework, we achieved coherent control over the nonlinear emission by an individual plasmonic nanoantenna[33] and demonstrated all-optical routing of the upconverted light by all-dielectric metasurfaces[34], with efficiency larger than 90%. Such an efficient routing mechanism is enabled by an internal parity conversion between the photons of the two underlying nonlinear processes, namely sum-frequency generation (SFG) and third-harmonic generation (THG). Here, by leveraging the same parity conversion, we experimentally demonstrate all-optical polarization modulation and steering of upconverted light using nonlinear interferometry. As in Ref 34, we simultaneously pumped a nonlinear metasurface based on AlGaAs nanopillars with a pulse at $\omega$ and its frequency-doubled replica at $2\omega$. In addition, we engineered the metasurface pitch so that the first diffraction order overlaps with specific points in the Fourier plane where the electric field polarizations of the light emitted via either the SFG or the THG process are orthogonal. In this way, by controlling the relative phase and polarization between the pump beams, we are able to fashion the polarization state of the upconverted signal at will in the first diffraction orders of the metasurface, routing opposite polarizations among the two axially symmetric orders. Toggling the relative phase delay between the pumps by $\pi$ we can completely swap the polarization state between the orders, obtaining circularly polarized light with a degree of circular polarization (DOCP) larger than 80%. Moreover, the polarization-modulated diffraction orders can be switched between optical planes by changing the polarization of the pump beams. Our work adds an additional layer of modulation – namely polarization – beyond intensity in an all-optical routing system implementing aprecise phase control. This capability to manipulate and route in $k$-space different polarization states, and in particular circular polarization states, discloses a significant potential for applications in chiral sensing, quantum computing and advanced imaging techniques.

**RESULTS**

The polarization modulation and routing of upconverted light is mediated by an all-dielectric AlGaAs metasurface. AlGaAs is particularly advantageous for photonic applications due to its high refractive index, which enables strong field enhancement crucial for nonlinear optical interactions. Its broad transparency window from the near-infrared (NIR) to mid-infrared (MIR) regions makes it versatile for a wide range of optical technologies. Additionally, AlGaAs features only off-diagonal components of second-order nonlinear susceptibility ($\chi^2$) with notably high value ($d_{36} > 100$ pm/V), resulting in high nonlinear conversion efficiency[35-39]. The nanopillars height is 400 nm and their radius is 250 nm, while the metasurface periodicity is fixed at 1000 nm to maximize the polarization mixing in the diffraction orther (see below).

The details on the fabrication can be found in the Materials Section and in Ref. 34. Figure 1a presents the concept of polarization modulation via nonlinear interferometry: the investigated metasurface enables interaction between a fundamental femtosecond pulse $\omega$ and its frequency-doubled replica $2\omega$, allowing precise control over the phase difference of the upconverted signals at $3\omega$ generated by either SFG or THG. The simulated (Comsol Multiphysics) SFG and THG far-field projections generated by each meta atom byhorizontally co-polarized pump beams at $\lambda_\omega$ = 1550 nm and $\lambda_{2\omega}$ = 775 nm are shown in panels b and c of Figure 1 and reveal that, because of the intrinsic opposite parities between $\chi^{(2)}$-mediated and $\chi^{(3)}$-mediated processes, the electric field distributions feature orthogonal electric fields for specific points in the *k*-space. We tailor the metasurface periodicity *p* to 1000 nm to have the (0, ±1) diffraction orders of the metasurface (see circles in panel b and c) to specifically coincide with these hot spots in the Fourier plane (see Ref. 34 for the mothods to retrieve the simulated back-focal plane maps). In this way, by controlling the relative phase delay between the two pump beams at $\omega$ and $2\omega$, we can gain full control over the polarization states of the upconverted light in these two diffraction orders. Figure 1d illustrates this concept: the polarization at $3\omega$ can be switched between circular and linear by varying the relative phase delay. Note that opposite polarization states, indicated in magenta and violet in panel d of Figure 1, are obtained in opposite locations in *k*-space with respect to the symmetry axis parallel to the $2\omega$ pump beam polarization. This configuration, which results from the internal parity conversion between the two different nonlinear optical processes, potentially enables to route different polarization states with opposite handedness between diffraction orders across the optical axis. By toggling the relative phase delay between the two processes by π, it is also possible to fully swap the polarization state between the orders.

To verify these predictions, we first run a full delay trace measurements using a mechanical stage place on the $\omega$ beam path and detecting the intensity of the upconverted light (SFG+THG) similarly to what reported in Ref. 34. In particular, we perform the Fourier imaging of the light upconverted by the metasurface by projecting the back aperture of the collection objective onto a CCD camera through 4*f* telescope without introducing an analyzer in the detection path (see Figure 2). This allows identifying specific diffraction orders reveal where the SFG and THG field intensities are orthogonally polarized, which would result in the absence of interference[34]. The polarization of the two pump beams is set to be horizontal with respect to the laboratory reference (*x* axis) as in the simulations of Figure 1b and c. A detailed description of the experimental setup is presented in the Materials and Method section. Figure 2a presents the delay-averaged back focal plane image of the upconverted light at $3\omega$. This image is obtained by an average of 40 frames captured at 1 fs intervals near the arbitrary zero-delay condition. In this case, the pump powers are $P_\omega = 43$ mW and $P_{2\omega} = 32$ $\mu$W to equalize the emitted SFG and THG powers. The diffraction orders $(0, +1)$, $(0, -1)$, $(-1, 0)$ and $(+1, 0)$ are anticipated at a numerical aperture (NA) of 0.52, where NA $= \lambda_{\text{THG/SFG}}/p$. Figs. 2b-i illustrate the full delay traces for all the diffraction orders within the collection NA (0.85) spanning the range from -200 fs to 200 fs, acquired with a resolution of about 1 fs. The intensity for each diffraction order is calculated by averaging a 13×13 pixel area around the peak and applying a background subtraction. Each delay trace displays a baseline of THG when the delay between the $\omega$ and $2\omega$ pulse is larger than the pulse width, while interference emerges when the former is lower than the latter, with the maximum interference identifying the zero-delay condition. The diffraction orders exhibit prominent interference fringes (blue traces) in Fig. 2b, d-g and i between SFG and THG. This is in line with our recent finding[34], and it is attributed to the internal parity conversion between the photons of the two nonlinear processes. In contrast, Fig.2c and h (red squares) display negligible interference fringes between SFG and THG in the (0, ±1) diffraction orders, pointing towards orthogonal field polarizations between the two processes. These two diffraction orders are thus particularly interesting for achieving tunable polarization states through phase modulation. To validate this evidence, we analyzed the polarization of the light emitted at $3\omega$ into these orders by inserting a linear polarizer in the detection path. This allows us to disentangle the contribution of the upconverted signals either along the $x$ or the $y$ axis and, hence, to clearly identify if the two frequency-degenerate signals possess orthogonal linear polarizations. Figure 3a and 3b show the delay traces with polarization selection along the $x$ (cyan traces) and the $y$ axis (blue traces) for the two diffraction orders exhibiting negligible interference (e.g. (0, +1) and (0, -1)). Both diffraction orders display a delay-independent baseline when the polarizer is aligned along the *x* axis, which is specific to a frequency tripling process only due to the $\omega$ beam that we attribute to THG. Concurrently, the delay traces show a bell-shaped profile when the analyzer is aligned along the *y* axis, indicating an upconversion process produced by the superposition of the beams $\omega$ and $2\omega$, which we therefore attribute to SFG. The small residual fringes arise from deviations from a perfectly orthogonal alignment between the polarizations of the SFG and THG fields. The THG and SFG intensities at the zero-delay condition have been equalized to enable polarization modulation between opposite circular polarization states. This condition has been achieved by setting the impinging powers of the $\omega$ and $2\omega$ pump beams to be $P_\omega$ = 43 mW and $P_{2\omega}$ = 32 $\mu$W, respectively. This feature is fundamental to realize polarization modulation in these specific diffraction orders, as shown below.

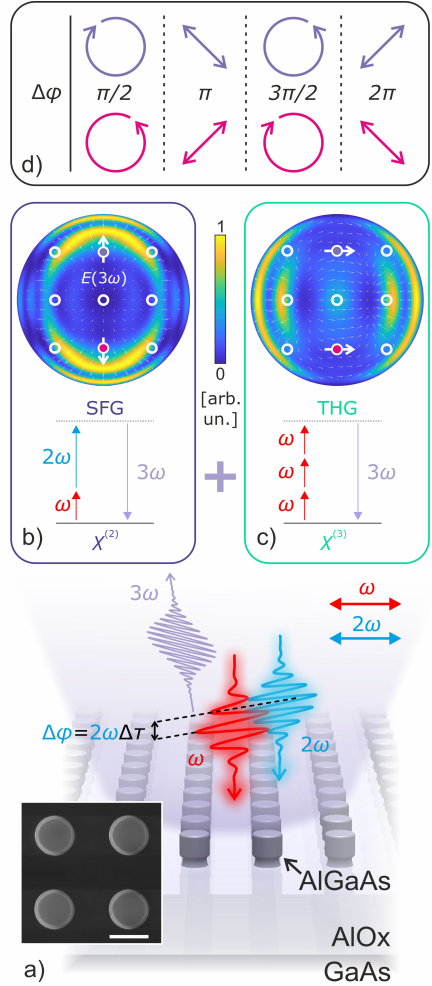

**Figure 1. All-optical polarization modulation and routing.** (a) Illustration of the nonlinear upconversion between a pulse at frequency $\omega$ and its frequency-doubled replica at $2\omega$, co-polarized and collinear, mediated by an AlGaAs metasurface. The inset shows the scanning electron micrograph of the investigated metasurface. Scale bar: 500 nm. Far-field angular distribution of the power emitted by each meta-atom via (b) sum frequency generation (SFG) and (c) third harmonic generation (THG). The white arrows indicate the direction of the electric fields at $3\omega$ in the $(0, \pm 1)$ diffraction orders of the metasurface. (d) Schematics of polarization control as a function of the phase delay $\Delta\varphi$ between the $\omega$ and $2\omega$ pump beams. Opposite polarization states are generated in the $(0, +1)$ and $(0, -1)$ diffraction orders (magenta and violet colors, respectively), that are in opposite locations with respect to the symmetry axis in the $k$-space parallel to the $2\omega$ pump beam polarization (see text for details).

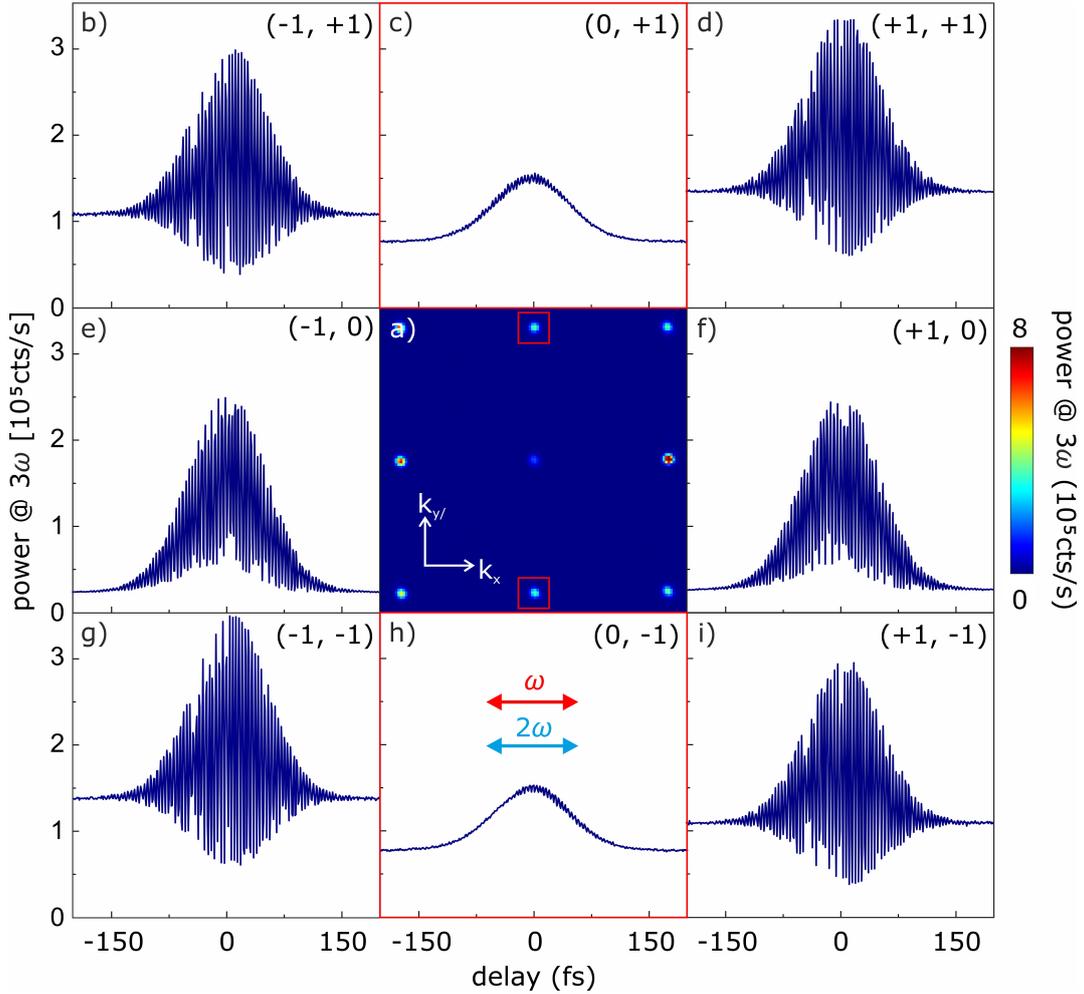

**Figure 2.** (a) Experimentally delay-averaged BFP image showing the upconverted light power at $3\omega$ (SFG + THG) from an AlGaAs metasurface with a lattice periodicity of 1000 nm. The image is produced by averaging 40 frames, each taken with a relative delay of 1 fs between the pump pulses at $\omega$ and $2\omega$. The linear polarization of the pump beams is aligned parallel to the $x$ axis, with powers of $P_\omega$ = 43 mW and $P_{2\omega}$ = 32 $\mu$W, respectively. The intensity of the light emission at $3\omega$ is measured by a CCD camera. (b-i) Full delay traces of eight diffraction orders. The intensity of the diffraction spots at each phase delay is obtained by integrating a 13×13 pixel area around the center of the diffraction spots and subtracting the background.

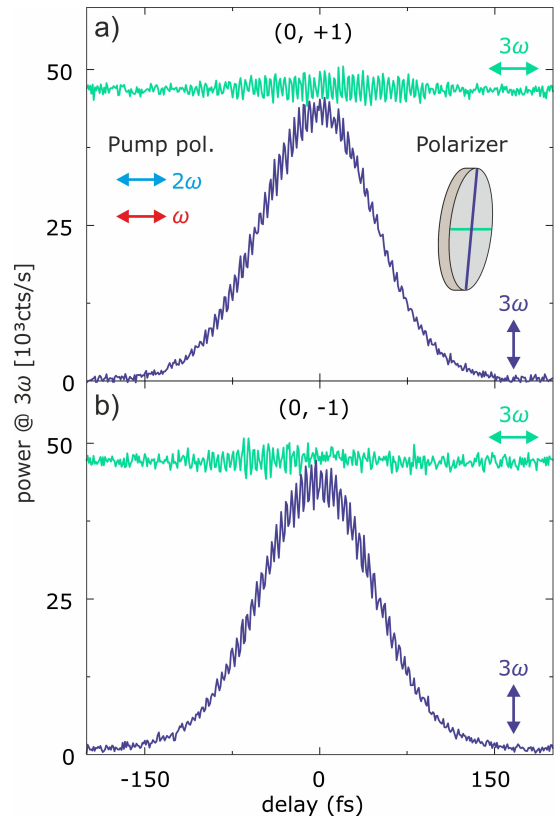

**Figure 3.** (a) Full delay traces of the (0, +1) diffraction order at $3\omega$, analyzed with a linear polarizer selecting the light linearly polarized along the $x$ (cyan) or $y$ (blue) axis in the detection path. When the linear polarizer is aligned along the $x$ axis, the trace shows a delay-independent behavior, to be ascribed to the THG component (cyan). In contrast, when the linear polarizer is aligned along the $y$ axis, the trace exhibits a Gaussian profile (blue), characteristic of the SFG component. The THG and SFG intensities at the zero-delay condition have been equalized to enable polarization modulation between opposite circular polarization states. The residual interference fringes on both delay traces are due to a non perfect orthogonality between SFG and THG polarizations. (b) Same as (a) for the (0, -1) diffraction order.

We highlight that by rotating the polaization of the $2\omega$ beam by 90°, hence by toggling the polaization configuration of the pumps from co-polarized to cross-polarized, such a configuration of the electric fields is routed from the (0, +1) and (0, -1) diffraction orders to the (-1, 0), (+1, 0) ones, i.e. from the vertical optical axis to the horizontal optical axis. This demonstrate the flexibility of this experimental realization that allows switching the polarization-modulated diffraction orders between different optical planes, offering an additional layer of modulation to the system.

To demonstrate polarization modulation, achieving a fine tuning of the $3\omega$ polarization state, we introduce a half-wave compensated liquid crystal retarder (LCR) in the $\omega$ beam path with its slow axis aligned parallel to beam polarization. By varying the applied voltage between 0 and 10 V, we can induce a relative phase delay ($\Delta\varphi = 2\omega \cdot \Delta\tau$) ranging from 0 to $2\pi$ with a resolution of $0.1\pi$ (150 as), corresponding to a full modulation cycle. To retrieve the complete polarization information of the upconverted light at $3\omega$, we employed the rotating quarter-waveplate polarimetry[40]. Briefly, this consists in inserting a rotating quarter-waveplate in the detection path followed by a fixed linear polarizer, in this case with the transmission axis aligned along *x*. Polarization modulation begins at an arbitrary zero-delay condition set by the different path lengths of the $\omega$ and $2\omega$ pump beams. The normalized Stokes parameters are evaluated over the full modulation cycle by measuring the diffraction order intensities from the BFP images acquired via the CCD camera. Data points are obtained by integrating a 13×13 pixel area around the center after background subtraction. Figure 4 show the normalized Stokes parameters as a function of delay for two different pump polarization configurations, co-

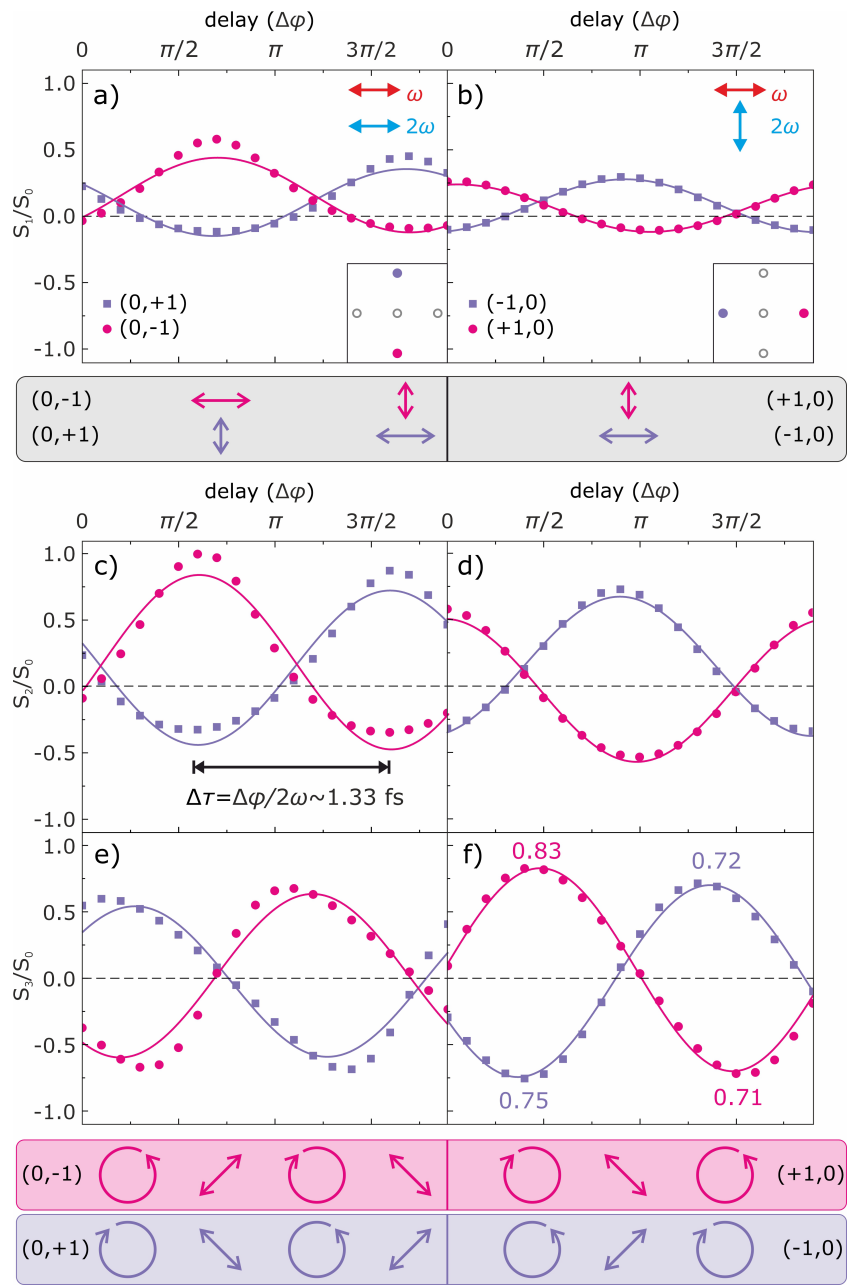

**Figure 4. Normalized Stokes parameters as functions of delay.** Normalized Stokes parameter $S_1/S_0$ for (a) the (0, +1) (squares) and (0, -1) (dots) diffraction orders when co-polarized pump beams aligned along the *x* axis are employed and (b) the (-1, 0) (squares) and (+1, 0) (dots) diffraction orders when cross-polarized pump beams are employed. The curves represent the corresponding sinusoidal fits. Inset in a and b: representation of the diffraction orders respectively investigated. The gray bar helps identifying the linear polarization states of the respective orders in panels a and b. Normalized Stokes parameter (c, d) $S_2/S_0$, and (e, f) $S_3/S_0$ for the same diffraction orders as in (a) and (b). The magenta and violet bars help identifying the modulation between linear and circular polarization states of the respective orders in panels c-f.

(panels a, c and e) and cross-polarized (panels b, d and f). The data points in Figure 4 represent the values obtained through the Stokes polarimetry analysis. The normalized Stokes parameters $S_1/S_0$ and $S_2/S_0$ together represent the degree of linear polarization (DOLP), whereas $S_3/S_0$ represents the degree of circular polarization (DOCP). By observing the sinusoidal behavior of the (0, +1) diffraction order one can notice that $S_1/S_0$ and $S_2/S_0$ oscillate in phase, while $S_3/S_0$ oscillates in antiphase. This indicates a periodic transition between linear and circular polarization states as the relative phase delay between two pump beams changes. Furthermore, the normalized Stokes parameters for the opposite (0, -1) and (+1, 0) diffraction orders exhibit antiphase behavior compared to those for the (0, +1) and (-1, 0) ones. Therefore, signals having opposite polarization states are symmetrically distributed in the Fourier space at each phase delay. In particular, circularly polarized light with opposite handedness and a high DOCP can be generated symmetrically across the optical axis at a specific relative phase delay. By introducing an additional π phase delay, the handedness of the circular polarization can be reversed. The polarization modulation mechanisms are schematized for clarity in the colored bars below panels a and b ($S_1/S_0$) and below panels e and f ($S_3/S_0$). Among all the configurations, we retrive a value of DOCP up to 83%. The limitation in achieving 100% DOCP arises from two primary reasons. First, the polarizations of SFG and THG fields are not perfectly orthogonal, and their powers are not perfectly balanced, which leads to elliptical polarization states and to a non-vanishing value of the $S_1/S_0$ ratio. Second, the DOCP is limited by the linear dichroism introduced by the optical elements, as confirmed by measuring the transmitted polarized light of a laser emitting around $3\omega$ (not shown here). These effects are also responsibe for the reduced amplitude and asymmetry of the $S_2/S_0$ and $S_3/S_0$ ratios, which indicate an unbalance between the +45° and −45° linearly polarized light (see Figure 4c and d) and between the left- and right-circularly polarized light (see Figure 4e and f).

**CONCLUSION**

In conclusion, we demonstrated the upconversion of telecom photons ($\omega$) to the visible range ($3\omega$) using a nonlinear AlGaAs metasurface. This was accomplished through two frequency-degenerate nonlinear optical processes: sum-frequency generation (SFG) and third-harmonic generation (THG), driven by a dual-beam pumping scheme ($\omega + 2\omega$). By analyzing individual diffraction orders in the Fourier space, we broke the mirror symmetry to generate chiral outputs in specific diffraction orders, although the system itself is achiral. This allowed us to modulate the polarization of the upconverted light at $3\omega$ between linear and circular states by controlling the relative phase delay between the pump beams. We identified diffraction orders with orthogonal polarizations between SFG and THG and achieved precise polarization modulation, by balancing their powers and dephasing them with a resolution of $0.1\pi$ using a liquid crystal retarder. Our results demonstrated polarization modulation between linear and circular states, with circularly polarized light of opposite handedness symmetrically generated across the optical axis with a degree of circular polarization larger than 80%. Moreover, the polarization-modulated diffraction orders could also be switched by adjusting the relative polarization of the $\omega$ and $2\omega$ pump beams. This method presents an alternative approach to polarization modulation for all-optical routing, complementing intensity modulation. Our robust technique, based on nonlinear interferometry for the control of the polarization state, can be extended to other nonlinear platforms. It holds significant potential for applications in chiral sensing, quantum computing, and advanced imaging technologies.

**MATERIALS AND METHODS**

**Sample Fabrication**

The metasurface we use consists of $Al_{0.18}Ga_{0.82}As$ nanocylinders, each 400 nm in height and 500 nm in diameter, oriented along the [001] crystal axis. These nanocylinders are fabricated via electron-beam lithography and supported by a low-refractive index AlOx substrate (n = 1.6), resulting in a strong field confinement within the AlGaAs nanocylinders (n = 3.2)[36]. This specific composition is selected to shift the bandgap above 747 nm, which suppresses two-photon absorption (TPA) to further enhance nonlinear conversion efficiency[39]. In this configuration, each nanocylinder supports an electric dipole resonance at a pump wavelength of 1556 nm[34].

**Experimental setup**

The laser source generates pulses with a duration of 200 fs and a repetition rate of 80 MHz, centered at an angular frequency $\omega$, corresponding to a wavelength of 1556 nm. The beam is partially frequency doubled to $2\omega$ with a beta barium borate crystal (Eksma Optics, BBO-SHG@1554nm), producing a wavelength of 778 nm. The two pulses are separated through a short-pass dichroic mirror (Thorlabs, DMSP950), with the relative temporal delay controlled precisely by a mechanical delay stage (Physik Instrumente, M-404) in the $\omega$ beam path, allowing for adjustments with a minimum step of 0.66 fs. Subsequently, the $\omega$ and $2\omega$ beams are recombined with another short-pass dichroic mirror (Thorlabs, DMSP950) and focused onto the back focal plane of an air objective (Nikon, CFI Plan Fluor 60XC, NA = 0.85) using an achromatic lens doublet (Thorlabs, AC254-500-B). This configuration enables a nearly collimated illumination on the sample, resulting in a spot size of 20 $\mu$m. Linear polarizers (Thorlabs, WPH05M-1550 and WPH05M-808) are inserted in both $\omega$ and $2\omega$ beam paths, respectively, to adjust the polarization of the two beams. The nonlinear emission at $3\omega$ is collected by the same objective in an epi-configuration and separated from the excitation light by a long-pass dichroic mirror (Thorlabs, DMLP650). The emission is then filtered by spectral filters (Thorlabs, FESH1000 + FESH700 + FBH520-40) to select the wavelength centered around 518 nm ($3\omega$). Eventually, the back focal image of the objective at each phase delay is projected onto a silicon CCD camera (Andor, iKon-M DU934P-BV) with a 4f imaging system composed of a pair of achromatic doublets (Thorlabs, AC508-500-B). As an analyzer for the linear polarization detection experiment (see Figure 3) we inserted a linear polarizer (Thorlabs, LPVISB100-MP2) in the detection path. High resolution phase delay plots were obtained by inserting a half-wave compensated liquid crystal retarder (LCR, Thorlabs, LCC1411-C) centered at 1551 nm with its slow axis aligned parallel to the polarization of the $\omega$ beam. The maximum resolution in the phase delay is about $0.1\pi$, which is equivalent to a temporal delay of 150 attoseconds[34]. For the Rotating Quarter-waveplate Polarimetry a quarter-waveplate (Thorlabs, AQWP10M-580) is rotated by an angle $\theta$, followed by a fixed linear polarizer (Thorlabs, LPVISB100-MP2) with its transmission axis aligned along the x-axis.